\documentclass[12pt]{iopart}
\usepackage{color}
\usepackage{graphicx}
\usepackage{epstopdf}
\usepackage{textcomp}
\definecolor{darkblue}{rgb}{0,0,.8}
\usepackage{bm}

\begin{document}
\title[Ions in MOT/BEC photonization] {Ion detection in the photoionization of  a Rb Bose-Einstein condensate}
\author{M. Viteau$\dag$, J. Radogostowicz$\dag$, A. Chotia$^+$, M. G. Bason$^+$, N.~Malossi$\dag$, F. Fuso$\dag$, D. Ciampini$\dag$, O. Morsch$^+$, I.~I.~Ryabtsev$^+$\footnote[3]{Permanent address:
Institute of Semiconductor Physics,  Pr. Lavrentyeva 13, 630090 Novosibirsk, Russia},  E.
Arimondo$\dag.^{+}$\footnote[4]{To whom correspondence should be
addressed (bason@df.unipi.it)}}

\address{$\dag$ CNISM UdR Pisa, Dipartimento di Fisica {\it E. Fermi},
Universit\`{a} di Pisa, Largo Pontecorvo 3, I-56127 Pisa, Italy}

\address{$^{+}$ CNR-INO, Dipartimento di Fisica {\it E. Fermi},
Universit\`{a} di Pisa,  Largo Pontecorvo 3, I-56127 Pisa, Italy}

 \date{\today}
\begin{abstract}
Two-photon ionization of Rubidium atoms in a magneto-optical trap and a Bose-Einstein condensate (BEC) is experimentally investigated. Using 100~ns laser pulses, we detect single ions photoionized from the condenstate with a 35(10)\% efficiency. The measurements are performed using a quartz cell with external electrodes, allowing large optical access for BECs and optical lattices.
\end{abstract}

\section{Introduction}
The development of laser cooling and trapping techniques has enabled the creation of dense and ultracold atomic samples, thus opening the way for laser spectroscopic investigations in regimes not accessible with conventional techniques. A wide range of applications are associated with ultracold atoms in different configurations, in particular the use of atoms in optical lattices and the configuration of a single atom per lattice site: the Mott insulator phase of bosonic atoms~\cite{bloch08}.  Atoms prepared in this way are candidates for qubit elements in quantum computation. Crucial issues to be solved in order to realize this quantum computation configuration are the initialization and readout operations on a single site of the lattice. In addition, high detection sensitivity will be required to isolate the signal of single atoms. 

Ion detection is a very efficient tool, because the charge amplification associated with existing devices increases the small signal generated by just a single ion to a measurable size. Laser photoionization can create ions or electrons with nearly unity probability using short and intense laser pulses.  Ion detection efficiency for single ion with channel electron multipliers (CEM) may reach up to 80-90 percent, provided the ion energy is high enough (3-5 keV). Thus, ion detection is an efficient tool for detecting a few atoms while not destroying the whole sample.
 
The investigation of photoionization of a cold cloud trapped in a MOT was first studied by Dinneen {\it et al}~\cite{dinneen92} using rubidium atoms, and later applied to other alkalis and alkaline-earth atoms. Using trap loss spectroscopy the two-photon ionization cross-section of rubidium ground state atoms in a MOT was measured~\cite{anderlini04,courtade04}. These works used the 6P$_{3/2}$ state as a near resonant intermediate state. Subsequent theoretical calculations of ionization cross-sections, stimulated by these cold atom measurements, are reported in references~\cite{anderlini04,potvliege06}. 

\indent In comparison with cold atoms in a MOT, the photoionization rate of a Bose-Einstein condensate (BEC) should be reduced by a Pauli blockade process because the single-energy products, ions and electrons, obey Fermi-Dirac statistics~\cite{mazets99}. However this predicted behavior is inhibited by the fast thermalization of electrons~\cite{killian}. Theoretical analyses predict that  the interactions between a BEC and a positively charged alkali-metal ion should lead to the rapid formation of mesoscopically large molecular ions and an excess number (positive or negative) of atoms around an ion created within the condensate~\cite{cote02,massignan05}. A comparison of photoionization losses from atoms in a MOT or a BEC confined by a magnetic trap indicated larger losses for the BEC~\cite{ciampini02}.

Rydberg-Rydberg collisions mediated by dipole-dipole interactions lie at the heart of several schemes of quantum information processing~\cite{jaksch00,lukin01}. Electron/ion collection is a basic technique for  the detection of Rydberg states created by exciting cold or ultracold atoms. Ion detection efficiency plays an important role in the spectroscopy of resonant collisions of a few Rydberg atoms, as discussed initially by Cubel Liebisch {\it et al}~\cite{cubel05} and later by Ryabtsev {\it et al}~\cite{ryabtsev07}. Such efficiencies are not always precisely known, making interpretation of experimental data more difficult. Detailed characterisations of ion detection efficiencies are thus worthwhile. As an example using cold atoms in a MOT, an estimation of electron detection efficiency yielded 1.5$\%$~\cite{teo03}.  Reference~\cite{ryabtsev07} describes a method to determine the ion detection efficiency and mean number of Rydberg atoms excited per laser pulse, and reports an initial efficiency of 13(1.5)$\%$, increased to 65(5)$\%$ by a detection system properly designed for the MOT~\cite{tretyakov09,ryabtsev09}.  More complex configurations, such as those with a double detector for electrons and ions have also been explored~\cite{loew07}. The research group of Zimmerman use Rb photoionization as a technique for the sensitive detection of atoms in the condensate~\cite{stibor07,kraft07}. For a MOT created on an atomic chip, the same research group recently reported single Rb atom counting rates of up to several MHz, with 67(12)$\%$ efficiency using three-photon ionization and subsequent ion detection with a CEM~\cite{guenther09}.  A high ion detection efficiency allowed the observation of sub-Poissonian statistics for ions produced by Rydberg excitation~\cite{cubel05}.

\indent This work examines rubidium BEC photoionization within an apparatus incorporating a quartz cell with large optical access. Such cells are useful for Rydberg atomic excitation and optical lattice confinement. As the electrodes are located outside the cell, some shielding of the electric field is seen. Care must also be taken not to charge the cell due to prolonged exposure to high voltages. Despite these limitations, collection and detection of ions is still efficient.
We report measurements of both atom number losses and ion generation resulting from photoionization. A systematic comparison of the photoionization products originating from the MOT or BEC allows the characterization of the processes of ion production and detection. From this analysis we deduce a large detection efficiency, $T$, in our apparatus of $35(10)\%$. Our results demonstrate that ion creation represents a powerful tool towards the target of single atom detection within a condensate.  

Section~\ref{apparatus} describes the experimental apparatus used to produce BECs, atomic ionization and charge collection.  The experimental results and their analyses for two different photoionization schemes are detailed in section~\ref{results}. Following the absorption of laser pulses with long, the ultracold atoms in the BEC recoil due to spontaneous emission. This effect, along with two-photon ionization spectra of the BEC, are presented. Calibration of the ion detection efficiency and the Poisson distribution associated with the ion collection process are also detailed. The photoionization rates of MOT atoms for two photoionization schemes are compared to those measured in a BEC.  Section 5 presents our conclusions.  An appendix recalls the theoretical approach introduced in reference~\cite{anderlini04} and used to extract the photoionization atomic parameters from the atom loss measurements. 

\section{Set-up}
\label{apparatus}
\indent The BEC apparatus uses two quartz cells, called collection and science cells respectively, horizontally mounted on a central metallic structure.  In the collection cell, $^{87}$Rb atoms are captured and cooled in a 2D MOT, producing a continuous flux of atoms from the collection to the science cell.  Within the science cell, the lifetime of the atoms in the 3D MOT is around two minutes.  The 3D MOT runs in two configurations to give different sizes, around 1~mm when the fluorescent signal is monitored and around 100--200~\textmu{}m during ion detection. A small MOT with a low number of cold atoms avoids saturation of the charge collection and allows uniform illumination of the atoms by the ionizing laser beams. To produce a BEC, atoms are first evaporatively cooled in a TOP magnetic trap, until the phase space density reach a value around 0.1. Then the atoms are loaded into an optical dipole trap created by two crossed beams derived from a 1030~nm Yb:YAG with 3~W cw power, focussed to a waist of 70~\textmu{}m. Using a PID lock to regulate the laser power ensures high stability of the potential in the dipole trap. Evaporative cooling in the dipole trap yields pure condensates of around $10^5$ atoms.  Absorption imaging after a 6.3~ms time-of-flight (TOF) is used to monitor the MOT/BEC before and after ionization.

\begin{figure}[ht]
\centering\begin{center}
\includegraphics{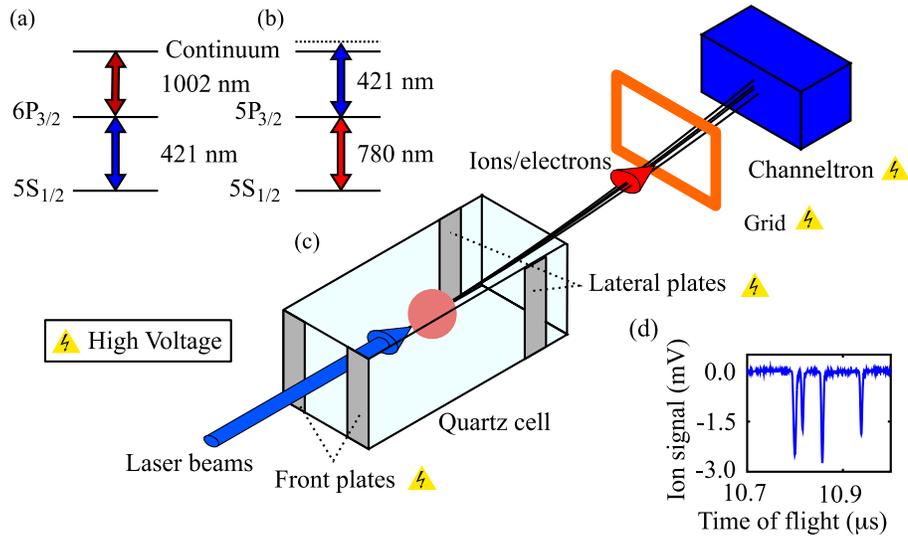}
\end{center}
\caption{In (a) and (b) the atomic levels involved in the two photoionization schemes are presented.
A schematic of the ion collection system produced by the MOT/BEC ionization within the vacuum quartz cell is shown in (c). The laser beams are superimposed so as to excite atoms within the same volume. The configuration with plates on the front and sides of the cell, grid and CEM to the rear is shown. Electrons or ions are collected depending on the signs of the applied voltages. In (d) a typical ion measurement shows the arrival of 4 ions .}
\label{ChargeCollection}
\end{figure}

 \indent To generate ions, two different schemes are used, as shown in figures~\ref{ChargeCollection}(a)~and~(b).  In the first scheme, the $\mathrm{6P}_{3/2}$ level lying 421~nm from the $\mathrm{5S_{1/2}}(F=2)$ ground state is used as a resonant intermediate state for two-photon absorption. The ionization threshold can thus be crossed from the $\mathrm{6P}$ state by absorbing a second 421~nm or a 1002~nm photon.  The second scheme is based on the atomic excitation to the $\mathrm{5P}_{3/2}$ state by the cooling lasers followed by absorption of 421~nm radiation, as shown in figure ~\ref{ChargeCollection}(b).  For this scheme in order to attain higher intensities, the 421~nm beam is focussed to a waist, $w$, of 70~\textmu{}m  larger than the BEC size. This arrangement maintains the BEC cloud within the beam waist in the presence of fluctuations in the spatial position of the condensate, less than 10~\textmu{}m. The ionizing laser beams are pulsed with duration $\tau_{\rm pul}$, between 0.3 and 1.2~\textmu{}s, with a gaussian rise time $\tau_{\rm rise}\sim$ 0.08~\textmu{}s
controlled by an acousto-optic modulator.  

By doubling a MOPA laser (TOPTICA TA 100, output power 700~mW) with a TOPTICA cavity, 60~mW of 421~nm radiation is generated. An infrared laser tuned between 1000 and 1030~nm allows ionization or spectroscopic resolved excitation to states from $n=30$ quantum number  to the continuum.  A combination of power and stability of the infrared radiation is achieved by injecting a Sacher diode (output power 40~mW) into a Sacher TIGER laser (output power 250~mW) where the grating is replaced by a mirror. Both laser systems, having a line-width smaller than 1~MHz, are locked using a Fabry Perot interferometer where a 780~nm laser locked to the Rb resonance acts as a reference~\cite{zhao98,rossi02}.

The electron/ion detection scheme is shown in figure~\ref{ChargeCollection}(c).  A CEM is located around 15~cm from the cell centre, close to the exit of the graphite tube between the collection and science cells.  A grid at a high voltage and located 10~cm from the ultracold atomic cloud guides the emitted ions/electrons.  High voltage plates placed outside the cell on the front and to the side guide the charges produced at the centre of the science cell to the CEM.  The ions are collected through a sequence of short electric pulses which typically have a rise time of a few ns. In the initial configuration, the high voltages were applied only to the front plates (3~kV for 5~\textmu{}s) and the grid (-1~kV for 25~\textmu{}s).\footnote{Rydberg spectroscopy indicates that the electric field acting on the atoms is roughly ten times smaller than that applied to the outside of the cell. This is due to partial shielding by the 2~mm thick, quartz walls.} To improve the detection efficiency a +3~kV pulse was applied to the front plates for 5~\textmu{}s and simultaneously a -1.5~kV pulse was applied to the lateral plates. Afterwards, a -1.2~kV pulse was applied to the grid. The length of all the applied electric fields are kept short to avoid charging the quartz cell. This configuration ensures a more efficient attraction of positive ions toward the CEM.  Ion collection was chosen in order to realize a temporal separation between the collected charge signal and  the spurious signal induced by the electric field switch on the CEM output.  The detection efficiency, $T$, represents a global efficiency taking into account both the fraction of produced charges transported to the detector and the CEM conversion efficiency. Its value, under both field arrangements, is reported in section~4. The sensitivity of the apparatus allows the monitoring of single ion signals, as can be seen in figure~\ref{ChargeCollection}(d). Depending on the number of produced ions and their arrival time spread, ion counting or signal integration by a boxcar is used.  From the histogram of the collected ion pulse areas, a conversion between output signal and that due to a single ion is derived.

\begin{figure}[ht]
\centering\begin{center}
\includegraphics{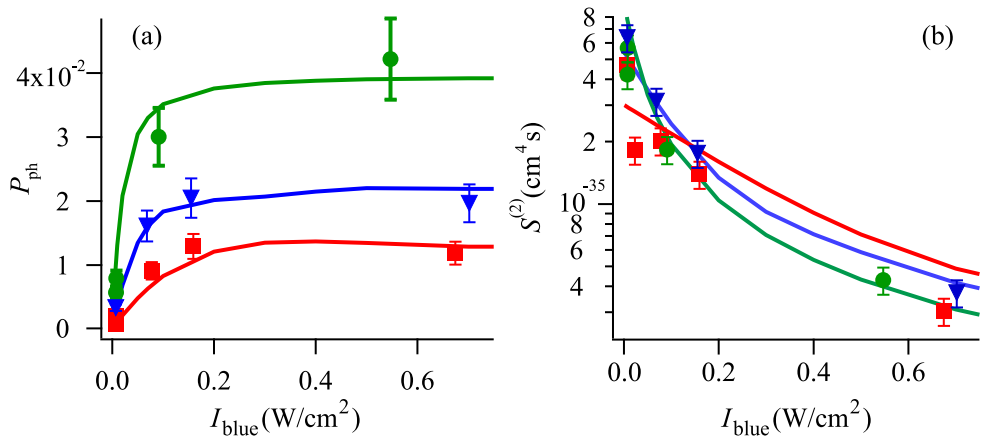}
\end{center}
\caption{(In colour on line) In (a), the $P_{\rm ph}$  photoionization probability per laser pulse and in (b), the $\cal{S}$$^{(2)}$ unitary flux photoionization rate   versus the  blue laser intensity $I_{\rm blue}$ at $I_{\rm ir}= 2000$~W/cm$^2$, and different pulse lengths, green circles $\tau_{\rm pulse}$ = 1~\textmu{}s, blue triangles 0.5~\textmu{}s and red squares 0.25~\textmu{}s,  having 80~ns rise and fall times. The continuous lines represent theoretical predictions for those different pulse lengths, green, blue and red, respectively.}
\label{Ionfraction}
\end{figure} 

\section{Results}\label{results}
\subsection{Two-photon MOT population losses}\label{421_1002_MOT}
Initially the 421+1002 nm ionization of the MOT atoms was measured through atom losses, as described in appendix A. The photoionization probability per pulse, $P_{\rm ph}$, and the unitary flux photoionization rate, ${\cal{S}}^{(2)}$ introduced in that appendix are plotted as a function of $I_{\rm blue}$ and $\tau_{\rm pulse}$ in figure~\ref{Ionfraction}(a) and (b). Both plots are in good agreement with theory, confirming the excellent control over photoionization, and also the importance of the pulse rise/fall times. The influence of saturation and shifts on the intermediate level is manifested in the change of ${\cal{S}}^{(2)}$ with pulse length and laser intensity. 
 
\begin{figure}[ht]
\centering\begin{center}
\includegraphics{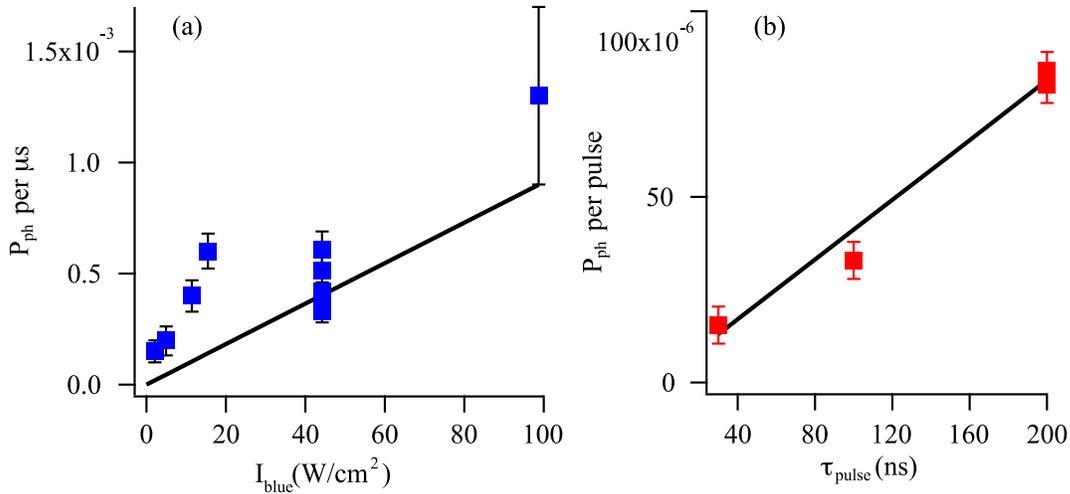}
\end{center}
\caption{In (a), 780 + 421 nm photoionization probability $P_{\rm ph}$ per \textmu{}s versus blue laser intensity $I_{\rm blue}$,
$\tau_{\rm pulse}$ =1~\textmu{}s. While (b) shows the photoionization probability per laser pulse, $P_{\rm ph}$ versus the pulse duration $\tau_{\rm pulse}$ for $I_{\rm blue}$ = 44.2~W/cm$^2$. These same data are reported on the left plot applying a linear temporal scaling. In both plots, the 780 nm laser intensity is 0.15 W/cm$^2$and has detuning -16.8~MHz from the $\mathrm{5S}_{1/2}$ $(F=2)$ $\to \mathrm{5P}_{3/2}$ $(F^\prime=3)$ resonance. The continuous lines represent theoretical predictions.}
\label{780421Ionfraction}
\end{figure}

MOT losses for the case of two-photon ionization via the 5P$_{3/2}$ level using the cooling laser at 780 nm and the 421 nm laser are presented in figure~\ref{780421Ionfraction}. For the range of intensities explored, $I_{\rm blue}$=~2-100 W/cm$^2$, the MOT ionization rate is strictly proportional to the product of the laser intensities.  Because the large laser intensity is applied to the transition from an excited state to the continuum, shifts and widths of the intermediate states and optical pumping were not important for the ionization process.  
The resultant photoionization probability is plotted in figure~\ref{780421Ionfraction}(a) versus the blue laser intensity and in figure~\ref{780421Ionfraction}(b) versus the pulse duration at fixed blue laser intensity, the continuous lines report the theoretical predictions. To properly analyse the MOT atom losses, the modification of the atomic cloud volume following ionization losses is taken into account. For this correction, it is assumed that the MOT operates in the constant density regime \cite{walker90}. 

\begin{figure}[ht]
\centering\begin{center}
\includegraphics[width=14cm]{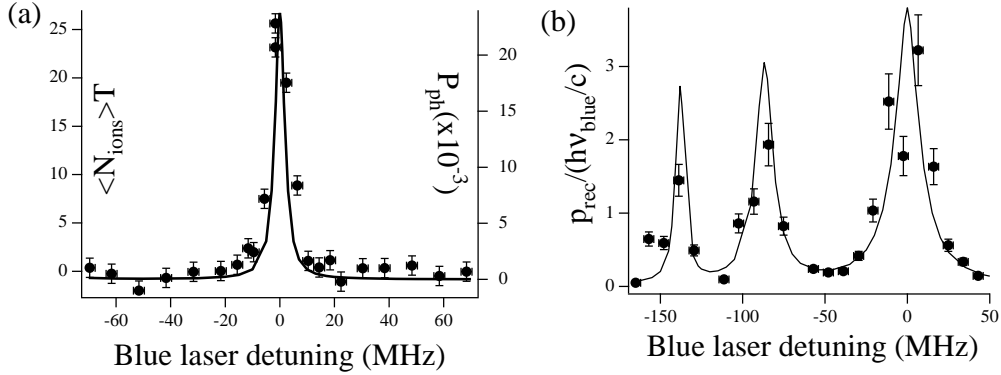}
\end{center}
\caption{In (a) ion spectra produced by BEC photoionization using 421+1002~nm laser beams versus blue laser detuning, at constant IR laser power are shown. The number of BEC atoms is around 1$\times$10$^5$. The number of collected ions is calculated from the output voltage, minus the average background, divided by the output voltage per ion as derived from the pulse histogram. Each data point is produced by a laser pulse with $I_{\rm blue}$=~0.12~W/cm$^2$. The continuous line shows the calculated photoionization probability $P_{\rm ph}$ (right scale).  Spectra observed on the  $p_{\rm rec}$ recoil of the ultracold cloud, derived from the displacement of the thermal cloud centre, versus the blue laser frequency at $I_{\rm blue}$~=~4~W/cm$^2$ are shown in (b). The continuous line shows the calculated recoil momentum $p_{\rm rec}$,  again measured in units of the photon momentum $h\nu_{\rm blue}/c$.  Zero detuning corresponds to the 5S$_{1/2}(F=2) \to$ 6P$_{3/2}(F^\prime = 3)$ transition. The resonant peaks correspond to the excitation of the upper $F^\prime$=1, 2, 3 hyperfine states. In both spectra $I_{\rm ir}$ = 2000~W/cm$^2$ and $\tau_{\rm pulse}=0.5$~\textmu{}s.}
\label{Spectra}
\end{figure}

\subsection{Ions produced from BEC}
\label{thermalcloud}
Both photoionization schemes are also investigated using BEC atoms. One target of the investigation was to verify whether or not photoionization of a BEC leads to different processes from those occurring in a MOT. By observing atom losses and ion creation, for the same laser parameters, no differences between the BEC and those measured on the cold atoms in the MOT are seen. However, the analysis of BEC atom losses should carefully consider the creation of thermal atoms following spontaneous emission from the excited state, as presented in section~\ref{thermalcloud}. 
 
Figure~\ref{Spectra}(a) shows the average collected ion number $<N_{\rm ions}>T$ in the BEC as a function of the 421~nm laser detuning with fixed 1002~nm laser wavelength.  Here, $<N_{\rm ions}>$ is the average number of ions produced. During this experiment, the integrated ion signal is measured: by comparing this number to that produced by a single ion, $<N_{\rm ions}>T$ is calibrated. These data are compared to the solutions of the time evolution of the density matrix under the action of two photoionizing laser fields. The theoretical photoionization probability $P_{\rm ph}$ per pulse is plotted as a continuous line in figure~\ref{Spectra}(a), with the right hand side vertical scale chosen to match the ion signal.  The comparison between experimental and theoretical results indicates that the blue laser frequency scanning introduces a 3~MHz broadening of the laser linewidth. The quantitative comparison between the theoretical predictions and the measured number of ions allows the derivation of the detection efficiency discussed in section~\ref{charge_collection}.


\begin{figure}[ht]
\centering\begin{center}
\includegraphics[width=8cm]{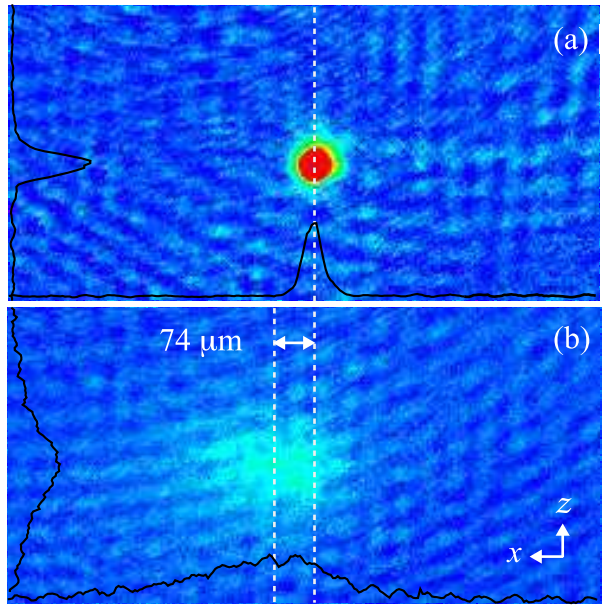}
\end{center}
\caption{In (a) an absorption image, after 6.3~ms time-of-flight, for a BEC cloud with initial trap dimensions of 14$\times$14~\textmu{}m is shown. The vertical lines indicate the centre of the unperturbed BEC and the centre of the displaced thermal cloud. An absorption image of the ultracold atomic cloud irradiated by a 0.3~\textmu{}s blue laser pulse with intensity $I_{\rm blue}= 0.7~$W/cm$^2$ is shown in (b). The $x$ and $z$ integrated profiles of the clouds are not shown to scale, but as a guide. The cloud displacement is due to the momentum imparted by blue photon absorption.  In addition, the spatial distribution of the ultracold atoms increases along the horizontal blue laser propagation direction, $x$, and also in the transverse direction $z$. }
\label{BEC_Recoil}
\end{figure}

\subsection{Photoinduced deformation of the BEC}
\indent The image of a BEC cloud before laser ionization shown in figure~\ref{BEC_Recoil} (a), reflects the momentum distribution of the BEC cloud. An image of the ultracold atomic cloud after a 0.3~\textmu{}s laser pulse composed of  421~+~1002~nm beams is shown in figure~\ref{BEC_Recoil}(b). Comparing the  absorption image with and without ionization indicates that 421~nm irradiation strongly modifies the ultracold atomic cloud. \\
\indent Following the absorption of blue laser running wave photons, a variable fraction of the condensate atoms recoil with momentum, $p_{\rm rec}$, and are subsequently converted into a thermal cloud. The centre of this thermal cloud is displaced along the propagation direction of the blue photons, the $x$ axis. Observing TOF images verifies that the momentum transfered by IR photons is negligible compared to that due to the blue photons. \\
\indent The density matrix equation model of photoionization mentioned in appendix is used to calculate the momentum transfered from the blue photons to the condensed atoms excited to the 6P$_{3/2}$ level.  For the data of figure~\ref{Spectra}(b), the calculated atomic cloud recoil is shown as a continuous line, in excellent agreement with the experimental results. The long lifetime of the 6P$_{3/2}$  excited state, compared to the 5P$_{3/2}$ state, limits the number of spontaneous emission processes and therefore the momentum transferred to the atoms.

\begin{figure}[ht]
\centering\begin{center}
\includegraphics{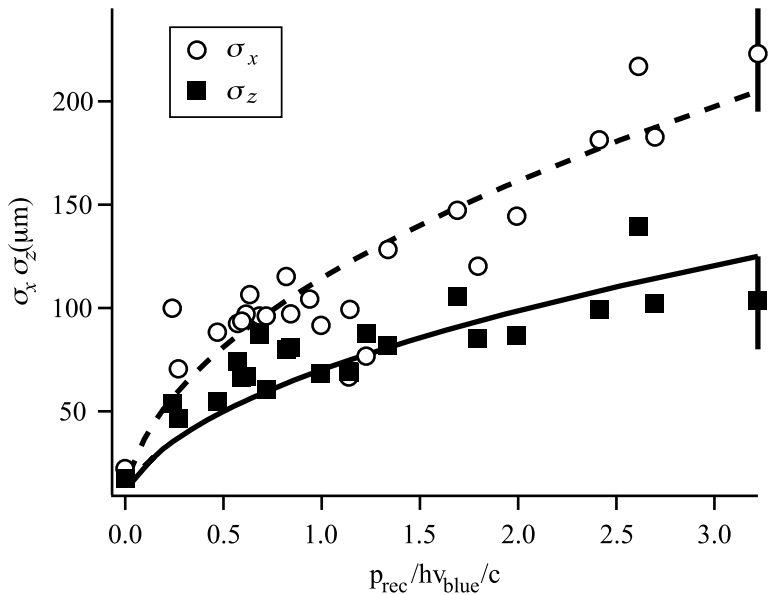}
\end{center}
\caption{Spatial widths $\sigma_{x}$ and $\sigma_{z}$ (in \textmu{}m) of the atomic thermal cloud created by the blue laser absorption versus the $p_{\rm rec}$ recoil imparted to the ultracold cloud. These data are derived from  the measurements of figure~\ref{Spectra}(b).  A typical error bar is shown. The widths are fitted by a square root dependence on $p_{\rm rec}$, including a convolution with the initial BEC spatial width.}
\label{BEC_Width}
\end{figure}

\indent The absorption of blue photons and the resultant spontaneous emission broaden the spatial distribution of the atomic cloud along the laser propagation direction and also orthogonally. This broadening is seen in figure~\ref{BEC_Recoil}(b). The atomic spatial distribution is composed of contributions from the broad thermal cloud and condensed fraction. The thermal cloud broadening is proportional to the square root of the atomic recoil, see figure~\ref{BEC_Width}. Such a spread in momentum, as expected for a random walk, is due to the spontaneous emission diffusion. The spread increases with the square root of the number of the spontaneous emission processes, as described by two-level laser cooling theory \cite{stenholm86,lett89}. Such treatment predicts equal diffusion in each direction, as we observed  following excitation of a condensate by 780~nm photons resonant with the 5P$_{3/2}$ state. In these measurements the momentum diffusion in perpendicular directions is in a ratio of $\approx\sqrt{3}$:1. Rather than a single decay channel, the 6P$_{3/2}$ decay paths include the 4D, 6S and 5P levels~\cite{courtade04}. The balance of spontaneous emission into different spatial directions thus governs the shape of the cloud. The exact role of the intermediate states would require an analysis of the generalised optical Bloch equations \cite{aspect89}. 
\subsection{Charge detection}\label{charge_collection} 
Given the determination of the MOT/BEC photoionization losses, the average number of ions collected by the CEM is measured as a function of 421~nm laser intensity and frequency. An example of these signals is given by the spectrum of figure~\ref{Spectra}(a). Such data allows the calibration of charge detection efficiency in the following manner. Using MOT depletion, the photoionization probability for a laser pulse is measured, as in Figure 2(a). Secondly, this probability is used to calculate the number of ions produced by a photoionization pulse with the parameters of figure~\ref{Spectra}(a). Finally, this number is compared to the number of ions detected by the CEM, $<N_{\rm ions}>T$, reported in figure~\ref{Spectra}(a), and thus $T$ is obtained. During this calibration the experiment operates at a low number of collected charges, which avoids saturating the CEM output. 

Ideally one would like to operate with large ionization losses in the MOT, so that the decay is measurably faster than the decay due to all other processes. This, however, would result in the production of a large number of ions to the point that the detection saturates as the setup is optimised for a low number of detected ions.

\begin{figure}[ht]
\centering\begin{center}
\includegraphics{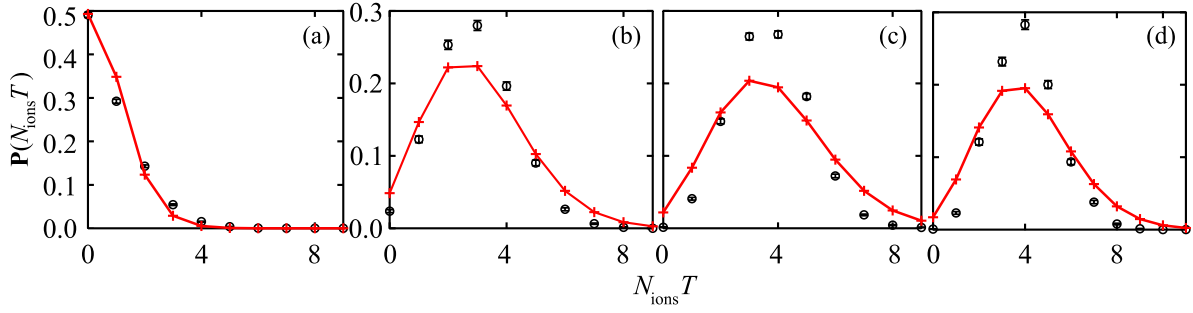}
\end{center}
\caption{Probability $\bf{P}$ of detecting $N_{\rm ions}T$ ions at the CEM output for different pulse durations, $\tau_{\rm pulse}$, in the 780 nm + 421 nm photoionization configuration. From (a) to (d) $\tau_{\rm pulse}=$ 30,~100,~150,~200~ns respectively. The 780~nm laser has intensity, 0.15~W/cm$^2$ and detuning --16.8~MHz while the 412~nm laser has an intensity of 45~W/cm$^2$. Lines connect the values of the Poissonian fit to the data. Measured  average value $<N_{\rm ions}>T$ of the collected ions  0.8(1), 2.9(3), 3.7(4) and 3.9(4), respectively from (a)--(d).  The corresponding values determined from a fit with a Poisson distribution are 0.70(4), 3.0(2), 3.8(3) and $4.1(3)$.}
\label{Poissons}
\end{figure}

The calibration of the average number of ions detected has been applied to the photoionization scheme 780~+~420~nm using a statistical analysis of the number distribution of the ion pulses at the CEM output (counted by a discriminator and a gated counter). The ion distribution shown in figure~\ref{Poissons}, where various pulse durations are used, is Poissonian, analogous to the ion distribution produced by Rydberg excitation~\cite{ryabtsev07}. The mean number $<N_{\rm ions}>T$ of ionization events extracted from these distributions was again compared the expected one, derived using the MOT decay measurements of Figure 3(a). This led to a value for the efficiency T in very good agreement with all different measurements.\\
\indent The spectrum of figure~\ref{Spectra}(a) was obtained using the configuration with electrodes on only the front of the science cell.  A maximum global detection efficiency of $T=3(1)\%$ was measured.  The Poisson distributions of figure~\ref{Poissons} were obtained in the configuration with electrodes on the front and side walls, resulting in an efficiency of $T=35(10)\%$. The BEC measurements demonstrated that $T$ is independent of the position of the degerate gas within the laser beam waist. 

\section{Conclusions} 
A new BEC experimental configuration for ionization and charge collection has been described and characterised.  Ionization investigations were facilitated by a 421~+~1002~nm excitation scheme via the 6P$_{3/2}$ level.  This laser combination also allows the excitation to Rydberg states by tuning the infrared laser to around 1020~nm. Later the 780~+~421~nm ionization scheme via the 5P$_{3/2}$ level was applied.  Ultracold atoms experience a recoil due to spontaneous emission from the excited level following photon absorption in the first step.  The recoil may be decreased by applying the ionization lasers in a standing wave configuration, using shorter duration laser pulses, or by detuning further from the intermediate resonant state. To compensate for a decrease in the ionization probability in these last two cases, a small beam waist, slightly larger than the BEC dimension was used.
Detuning from the intermediate state, ion production is close to being non-destructive. The losses from the atom cloud result from either ion creation or the scattering of photons. For a 1~\textmu{}s light pulse detuned +500~MHz from the 6P$_{3/2}$ state, the probability of losing one atom from a condensate of $\sim$10$^5$~atoms due to absorption of a photon is approximately 1\%.\\
\indent Ionization spectra produced from a BEC have been presented. Using electrodes external to a quartz cell, electric fields guide charges towards a CEM detector. This scheme has been tested for the generation and guiding of ions originating from Rydberg excitation. The signal to noise ratio is high enough to detect the signal produced by a single ion. The large detection efficiency allows the reliable observation of the signal produced by the ionization of as few as three atoms. A next-generation design of the set-up, with reduced CEM to BEC separation should further increase the collection efficiency. Using single ion detection with a photoionization probability $P_{\rm ph}$ equal to 1, possible with short and intense laser pulses, would allow the monitoring of the signal produced by a single atom in a single measurement. Combining single-site addressing techniques and the ion detection approach discussed should enable the readout of single atom states in optical lattices. We have not yet reached this high efficiency goal, but our progress represents an important step to realising such a target within a BEC apparatus where optical access for loading a BEC into an optical lattice is available. 

\section{Acknowledgments}
Financial support was provided by the European Commission through the NAME-QUAM STREP Project and the EMALI Network,  by the CNISM
 through ``Progetto Innesco 2007'',  by the MIUR through PRIN-2007 Program, by the Einstein Italy-Russia collaboration,  by the COGITO partnership between the
UniversitŽ Paris Sud and the University of Pisa. I.R. thanks the support by the Russian Foundation for Basic Research within the EINSTEIN Consortium-RFBR bilateral project ``Nonlinear dynamic resonances in collective interactions of cold atoms" The authors  gratefully acknowledge  the contribution by A.~Zenesini to the initial photoionization measurements and the remarks by P.~Lambropoulos on the photoionization analysis.  
 
\section{Appendix A: Cold atom photoionization}
The dynamics of the number of trapped atoms in a MOT, with the trap loading shut off and in the constant density regime, is described by an exponential decay with a time constant $1/\gamma$.  In the presence of ionization, additional losses accelerate the atom number decay. The evolution of the ground state atomic number, $N$, in the MOT is given by the following decay law \cite{dinneen92}:
\begin{equation}
\frac{dN}{dt}= -(\gamma +\Gamma) N~.
\label{eq:MOTdecay}
\end{equation}
Where the two-photon ionization rate $\Gamma$, produced by the 421~nm blue laser with frequency $\nu_{\rm blue}$, intensity $I_{\rm blue}$ and by the 1002~nm infrared laser with frequency $\nu_{\rm ir}$,
intensity $I_{\rm ir}$  can be written as
\begin{equation}
\Gamma= g^{(2)} {\cal{S}}^{( 2)}  \frac{I_{\rm
blue}}{h \nu_{\rm blue}}  \frac{I_{\rm ir}}{h \nu_{\rm ir}}~.
\label{crosssection}
\end{equation}
\indent ${\cal{S}}^{(2)}$, with dimensions L$^4$T, determining the two-photon ionization rate for unitary fluxes of the two lasers, is explained later.  The geometrical correction coefficient $g^{(2)}$ introduced in reference~\cite{ciampini02} represents the ratio between the volume excited by the photoionization lasers and the volume occupied by the atomic cloud. By describing the MOT spatial distribution as a Gaussian with dimensions $d_{\rm x}=d_{\rm y}$ and $d_{\rm z}$, $g^{(2)}$ is given by \cite{ciampini02}
\begin{equation}
 g^{(2)} =\frac{1}{\sqrt{1+2\left(\frac{d_{\rm x}}{w}\right)^2}{}\sqrt{1+2\left(\frac{d_{\rm y}}{w}\right)^2}}~.
\label{g2}
\end{equation}
At constant $g^{(2)}$,  Eq.  (\ref{crosssection}) leads to a dependence of the photoionization rate, $\Gamma$, on the product of the laser intensities. That dependence is valid only at weak laser intensities when the shifts and widths of the intermediate states near resonance for the two-photon process are not
important~\cite{lambropoulos74}. When these conditions are not met, $S^{(2)}$ is interpreted as a generalized two-photon cross-section. A formula similar to Eq. \ref{crosssection} describes the ionization produced by the 780 nm + 421 nm sequence via the 5P$_{3/2}$ level.

Without ground state depletion the photoionization probability per pulse  $P_{\rm ph}$, integrated over $\tau_{\rm pulse}$, the pulse duration, is given by $P_{\rm ph} \sim \tau_{\rm pulse} \Gamma$.  $P_{\rm ph}$ and/or the unitary flux photoionization rate ${\cal{S}}^{(2)}$, may be determined by measuring the temporal decay, $\gamma$, of the MOT number in the absence of photoionization and the $\gamma+\Gamma$ decay in the presence of photoionization. 

A theoretical determination of $P_{\rm ph}$ is obtained from the time evolution of the density matrix equations for atoms interacting with two photoionizing laser pulses. The theoretical model developed in reference~\cite{anderlini04} takes into account not only the three states directly coupled by laser radiation, but also all the atomic states involved in the spontaneous emission cascade from the 6P to the 5S state. In addition, that model considers optical pumping into the 5S$_{1/2}(F=1)$ state modifying the ionization probability for excitation nearly resonant with the intermediate state.  The present analysis is based on the ionization cross-sections determined in~\cite{anderlini04}, the key ones being 12.4~Mbarn for the 421~nm ionization of the 5P$_{3/2}$ level, 4.7~Mbarn for that of the 6P$_{3/2}$ level,   and 16.4~Mbarn for the 1002~nm ionization of the 6P$_{3/2}$ level.

\section*{References}
\bibliographystyle{apsrmp}

\begin{thebibliography}{0}
\expandafter\ifx\csname natexlab\endcsname\relax\def\natexlab#1{#1}\fi
\expandafter\ifx\csname bibnamefont\endcsname\relax
  \def\bibnamefont#1{#1}\fi
\expandafter\ifx\csname bibfnamefont\endcsname\relax
  \def\bibfnamefont#1{#1}\fi
\expandafter\ifx\csname citenamefont\endcsname\relax
  \def\citenamefont#1{#1}\fi
\expandafter\ifx\csname url\endcsname\relax
  \def\url#1{\texttt{#1}}\fi
\expandafter\ifx\csname urlprefix\endcsname\relax\def\urlprefix{URL }\fi
\providecommand{\bibinfo}[2]{#2}
\providecommand{\eprint}[2][]{\url{#2}}

\end{thebibliography}


\begin{thebibliography}{99}

\bibitem{bloch08}  I. Bloch, Nature  {\bf 453} 1016 (2008).

\bibitem{dinneen92} T. P.~Dinneen, C. D.~Wallace, K. N.~Tan, and P. L.~Gould,
Opt. Lett. {\bf 17}, 1706 (1992).









\bibitem{anderlini04} M.~ Anderlini, E.~ Courtade, D.~ Ciampini, J.H.~ M\"uller, O.~Morsch, and E.~ Arimondo,  J. Opt. Soc. Amer. B {\bf 21}, 480 (2004). 

\bibitem{courtade04} E Courtade, M Anderlini, D Ciampini, J H M\"uller, O Morsch, E Arimondo, M Aymar and E J Robinson,  J. Phys. B: At. Mol. Opt. Phys. {\bf 37},  967 (2004).



\bibitem{potvliege06} R.M.~Potvliege and C. S.~Adams, NJP {\bf 8}, 163 (2006).

\bibitem{mazets99}I.E.~Mazets, Quantum Semiclass. Opt. {\bf 10}, 675 (1999).

\bibitem{killian} T. C. Killian, T.~Pattard, T.~Pohl, and J.M.~Rost, Phys. Rep. {\bf  449}, 77 (2007).

\bibitem{cote02} R. C\^ot\'e, V. Kharchenko, and M. D. Lukin, Phys. Rev. Lett. {\bf 89}, 093001 (2002).

\bibitem{massignan05} P. Massignan, C. J. Pethick, and H. Smith, Phys. Rev. A {\bf 71}, 023606 (2005).

\bibitem{ciampini02} D.~Ciampini, M.~Anderlini, J. H.~M\"uller, F.~Fuso,
O.~Morsch, J. W.~Thomsen, and E.~Arimondo, Phys. Rev. A {\bf 66},
043409 (2002).

\bibitem{jaksch00} D. Jaksch, J. I. Cirac, and P. Zoller, S. L. Rolston, R. C\^ot\'e and M. D. Lukin, Phys. Rev. Lett. {\bf 85}, 2208 (2000).

\bibitem{lukin01} M. D. Lukin, M. Fleischhauer, and R. C\^ot\'e, L. M. Duan, D. Jaksch, J. I. Cirac, and P. Zoller
Phys. Rev. Lett. {\bf 87}, 037901 (2001).

\bibitem{cubel05} T. Cubel Liebisch, A. Reinhard, P. R. Berman, and G. Raithel, Phys. Rev. Lett. {\bf 95}, 253002 (2005),  {\it ibidem} {\bf 98}, 109903 (2007)

\bibitem{ryabtsev07} I. I. Ryabtsev, D. B. Tretyakov, I. I. Beterov, and V. M. Entin, Phys. Rev. A {\bf 76}, 012722 (2007); {\it ibidem} {\bf 76}, 049902E (2007).

\bibitem{teo03} B. K. Teo, D. Feldbaum, T. Cubel, J. R. Guest, P. R. Berman, and G. Raithel, Phys. Rev. A {\bf 68} 053407 (2003).

\bibitem{tretyakov09} D.B.Tretyakov, I.I.Beterov, V.M.Entin, I.I.Ryabtsev, P.L.Chapovsky,  JETP {\bf 108}, 347 (2009). 

\bibitem{ryabtsev09} I.I.Ryabtsev, D.B.Tretyakov, I.I.Beterov, V.M.Entin,  arXiv:0909.3239 (2009).

\bibitem{loew07} R. L\"ow, U. Raitzsch, R. Heidemann, V. Bendkowsky, B. Butscher, A. Grabowski, and T. Pfau, arXiv:0706.2639.

\bibitem{stibor07} A. Stibor, S. Kraft, T. Campey, D. Komma, A. G\"unther, J. Fort‡gh, C. J. Vale, H. Rubinsztein-Dunlop, and C. Zimmermann, Phys. Rev. A {\bf 76} 033614 (2007).

\bibitem{kraft07}  S. Kraft, A. G\"unther, J. Fort\'agh, and C. Zimmermann, Phys. Rev. A {\bf 75} 063605 (2007).

\bibitem{guenther09} A. G\"unther, H. Bender, A. Stibor, J. Fort\'agh, and C. Zimmermann, Phys. Rev. A {\bf 80} 011604 (2009). 







\bibitem{zhao98} W. Z. Zhao, J. E. Simsarian, L. A. Orozco, and G. D. Sprouse, Rev. Sci. Instrum. {\bf 69}, 3737 (1998).

\bibitem{rossi02} A. Rossi, V. Biancalana, B. Moi, and L. Tomassetti, Rev. Sci. Instrum. {\bf 73}, 2544 (2002).

\bibitem{walker90} T.~Walker and D.~Sesko and C. Wieman, Phys. Rev. Lett. {\bf 64}, 408 (1990).


\bibitem{stenholm86}  S. Stenholm, Rev. Mod. Phys. {\bf 58}, 699 (1986).

\bibitem{lett89} P. Lett, W.D. Phillips, S.L. Rolston, C.E. Tanner, R.N. Watts, and C.J. Westbrook, J. Opt. Soc. Am. B {\bf 6}, 2084 (1989).

\bibitem{aspect89} A. Aspect, E. Arimondo, R. Kaiser, N. Vansteenkiste and C. Cohen-Tannoudji, J. Opt. Soc. Am. B {\bf 6}, 2112 (1989).

\bibitem{lambropoulos74} P.~Lambropoulos, Phys. Rev. A {\bf 9}, 1992 (1974).



\end{thebibliography}

\end{document}